# Exploring the Impact of Cochlear Implant Stimulation Artefacts in EEG Recordings: Unveiling Potential Benefits

Hongmei Hu[*1,2]

Ben Williges[1,3]

Deborah Vickers[1]

[2]Department of Medical Physics and Cluster of Excellence "Hearing4all", University of Oldenburg, Germany
hh594@cam.ac.uk, hongmei.hu@uol.de

[3]ENT Department, University Hospital, Essen, Germany
ben.williges@uk-essen.de

[1]SOUND Lab, Hearing Group, Department of Clinical Neuroscience, Cambridge University, UK
dav1000@medschl.cam.ac.uk

*Abstract - Given rising numbers of bilateral cochlear implant (CI) users, predominantly children, there is a clinical need for efficient and reliable tests that can objectively evaluate binaural hearing. These tests are crucial for guiding the setup of bilateral CIs to optimise delivery of binaural cues. Our primary goal is to introduce a clinical electroencephalogram (EEG) procedure to assess binaural hearing function at various stages within the auditory pathway. Previous research demonstrated that bilateral CI users significantly decrease in ability to discriminate interaural time differences when pulse rates exceed 300 pulses per second. Our paradigm utilizes different pulse rates to objectively explore the limits. A notable challenge with this EEG procedure is the interference induced by CI electrical stimulus artefacts. Despite this obstacle, the potential benefits of CI stimulation artefacts often go unnoticed. This paper outlines positive applications of the frequently criticized CI artefacts for optimizing the experiment setup.*

## I. INTRODUCTION

As the number of bilateral cochlear implant (CI) users grows, particularly for infants and children who cannot respond reliably to environmental stimuli, there are clinical demands for time efficient and accurate tests to objectively assess binaural hearing and to inform bilateral CI fitting. The long-term objective of our ongoing project is to develop a clinical EEG procedure capable of assessing the neural encoding of spatial cues across individuals with diverse hearing profiles. Thus far, a binaural EEG paradigm [1] has been developed and tested on young individuals with normal hearing (NH). This paradigm was devised to record multiple EEG responses (subcortical auditory steady-state responses (ASSRs) and cortical auditory evoked responses (CAEPs)) within a single session, using different stimulation rates. CAEPs responses consist of stimulus onsets, offsets, and interaural time difference (ITD) changes [1], named ITD acoustic change complex (ACC) responses.

Results obtained from young NH listeners suggest that ACC responses to fine structure ITD ($ITD_{FS}$) changes hold potential as an objective tool for assessing binaural sensitivity. However, the filtered click stimuli (representative of CI stimuli) evoked no detectable or much smaller ACC amplitudes for ITD envelope ($ITD_{ENV}$) changes compared to $ITD_{FS}$ changes. The same stimuli also elicited smaller ASSRs than for sinusoidally amplitude modulated (SAM) tones for carrier frequencies below 1600 Hz.

Bilateral CI users have demonstrated sensitivity to ITD when stimulated with single or multiple electrodes. However, there are substantial differences in performance between bilateral CI users and NH listeners in various binaural tasks (e.g., [2-4]). For example, CI users have about 5-10 times higher ITD detection thresholds compared to NH listeners, especially at higher pulse rates. ITD sensitivity for bilateral CI users at pulse rates above 300 pulses per second (pps) declines (e.g., [5, 6], see review [7]), very similar to the $ITD_{ENV}$ sensitivity of NH listeners (e.g., [8-11]). Various animal and computer models (e.g., [12-16]) have been used to understand the mechanisms underlying this rate-dependent degradation in bilateral CI users. However, the dependence of ITD sensitivity on pulse rate has not been systematically quantified using electroencephalogram (EEG) measures in CI listeners.

This study aims to optimize the aforementioned EEG paradigm [1] for bilateral CI users, employing direct stimulation or research CI processors. In CI stimulation-evoked EEG recordings CI stimulation artefacts are commonly found (as discussed in, e.g., [17-26]), particularly in the context of electrically evoked ASSRs (eASSRs).

Rather than focusing on reducing CI stimulation artefacts, this paper uniquely demonstrates how to use these often-criticized CI stimulation artefacts to improve laboratory setups and experimental designs for bilateral CI stimulation in clinically applicable EEG measurements. This paper aims to address certain frequently posed but not publicly documented queries. For instance:

Can multiple physiological responses used in [1] be recorded from bilateral CI users within a restricted time frame (<1 hour)? Are there embedded artefacts in the system that could potentially mask the real neural responses? Could unintended jitters be introduced in the

This work was mainly supported by Medical Research Council grant MR/S002537/1 and partially by German Research Foundation DFG/EMSATON/415895050.

stimuli? Can satisfactory results be obtained when using auto power-up frames in bilateral CI stimulation for Cochlear CI users? How can the recorded CI stimulation artefacts be employed to optimize clinic CI-EEG experiment setups?

Our aim is to shed light on these inquiries and facilitate a more comprehensive understanding of the potential complexities associated with EEG recordings in the context of bilateral CI users.

## II. METHODS

### A. Procedure

A series of baseline measures to profile participants were conducted including completion of the short version of the Speech, Spatial, and Qualities of Hearing Questionnaire (SSQ12) [27], speech-in-noise test, loudness scaling, left/right discrimination ability and centralization to assess how balanced hearing was across ears. Initial decive checks (electrode contact impedances, compliances, clinical CI Maps) were conducted to ensure safe stimulation levels and identify deactivated electrodes. Impedances were re-checked at the end of each appointment. For the tasks involving loudness scaling, centralization, and left/right discrimination, the electrode pair (left and right) closest to the 1-kHz centre frequency was chosen based on the clinic frequency allocation table.

Considering the loudness summation [28], a comfortable but soft level (scale 5) was selected monaurally using a 10-point loudness scale chart, aiming at bilateral loudness between scale 6 (the most comfortable level) and scale 7 (loud but comfortable) when stimulated binaurally. Adjustments were made to the left and right levels during the centralization procedure if the image was not centralized, with the sum of left and right levels remained constant.

Following centralization procedure, the ITD sensitivity was examined in a similar way to [18, 19, 29], for each pulse rate at a specific ITD (e.g., 1000 μs): Two consecutive intervals were presented on each trial, separated by 200 ms. Each interval contained four consecutive 400-ms unmodulated biphasic pulse trains (including 20-ms raised cosine rise/fall ramps), separated by 100 ms. In one interval, chosen at random, the four 400-ms pulse trains were the same, with ITD of 0 (e.g., A-A-A-A). In the other interval, the first and third pulse trains were the same as in the first interval, while the second and fourth pulse trains had a specific none-zero ITD (e.g., A-B-A-B). For instance, in a trial, a sequence could be A-B-A-B (interval 1), and A-A-A-A (interval 2). Participants were asked to indicate which of the two intervals contained a sequence that gave the perception of moving within the head. A brief training session was conducted prior to the main experiment to ensure participants understood the task.

### B. Participant

Due to article length restrictions, an exemplary dataset from one sequentially implanted bilateral CI participant (CI1, female, age 49) is reported in the results. CI1 exhibited notably good binaural hearing abilities for a CI user, as indicated by both SSQ12 and the left/right discrimination tasks. Participant provided voluntary written informed consent and was compensated with hourly pay for their participation, with the approval of the Ethics Committee of the University of Cambridge (PRE.2019.093).

### C. EEG Stimuli

The stimulus used was a sequence of unmodulated charge-balanced biphasic pulses, with 25-μs phase duration, and 8-μs interphase gap. This stimulus was repetitively presented to the fixed electrode pair, e.g., left and right CI electrodes 11 for CI1, at five pulse rates (40, 80, 160, 320, and 640 pps), using monopolar MP1+2 stimulation mode. The duration of each presentation is 12 s (Fig. 1 and 2). The stimulus is an ABACAS sequence. It includes 2 s of the diotic stimuli (A, ITD = 0, T1), following 2 s of the dichotic stimuli (B, ITD = 1000 μs, T2), then again 2 s of the standard stimuli (A, ITD = 0, T3), following 2 s of the dichotic stimuli (C, ITD = -1000 μs, T4), following 2 s of the diotic stimuli (A, ITD = 0, T5), and 2 s of silence (S). For every pulse rate, 30 repetitions were collected in each session. Only results of the first three pulse rates (40, 80, and 160 pps) are shown in this paper.

### D. Apparatus

The stimuli were controlled through a stimulation PC running MATLAB, which interfaced with the Nucleus Implant Controller 4.1 (NIC 4.1) via two PODs. These clinically used PODs connected the NIC 4.1 programming software to two CP910 off-the-shelf sound processors (Cochlear Limited, Sydney, Australia). Hardware clocks of the processors were synchronized using a reference tone played back with a FireFace UCX II sound card connected to the processors via a Cochlear Nucleus Bilateral Personal Audio Cable.

For most psychoacoustic with direct CI stimulation and EEG tests, the Oldenburg AFC framework for MATLAB [30] was used. Prior to the experiment, the stimuli were verified using two detector boxes and an oscilloscope.

EEG recordings were acquired using a high resolution BioSemi ActiveTwo system (Amsterdam, The Netherlands) with 64 channels, following the international 10–20 system. The sampling rate was set at 16384 Hz and each sample had a resolution of 24 bits. Supplementary channels were positioned on the left and right mastoids, while eye movements were captured using channels placed at the left infra-orbital and right lateral canthus locations. Voltage offsets consistently remained below ±40 mV, typically within ±20 mV. Scalp channels around the CI coil were unconnected. Four EEG recording electrodes with reference electrode Cz as used in [1, 18, 19] were of primary interest in this paper: right and left mastoids, Inion, and the channel approximately 3.5 cm below the Inion. A trigger signal, transmitted from one POD and elongated by a pulse stretcher, connected to the BioSemi system's trigger input.

Participants were seated in a recliner within an electrically shielded, sound-attenuating booth and their behavioural responses were entered by one of the authors into the response interface. During the EEG experiments, participants watched silent, subtitled movies and were instructed to minimize movements and disregard the stimuli.

**E. EEG analysis**

Continuous EEG data were segmented into epochs spanning a 13-s window, including a 400-ms pre-stimulus period. Following segmentation, the data were averaged across epochs and digitally filtered using a two-order Butterworth band-pass filter between 0.1 and 1500 Hz without applying any advanced CI artefact reduction. Baseline was defined by utilizing the mean amplitude of the 200 pre-stimulus timeframe, and no thresholding procedure was employed. The EEG data were then averaged across the trials independently for each condition. Same as depicted in [1], the responses were the voltage differences between the channel Cz and the average of the four clinical recording channels.

To explore the time-frequency characteristics of the evoked responses for different pulse rates, an adaptive super-resolution wavelet transform described in [31] was applied to the band-pass filtered signal (ranging from 0.1 to 1500 Hz). The resulting scalograms, presented in arbitrary unit (a.u.) were shown. For an in-depth comprehension of the scalograms, refer to Supplementary Information II in [31].

Consistent with [1], to obtain the eASSRs in the frequency domain, the band-pass filtered data between 0.1 and 1000 Hz were used. EASSRs were extracted within the whole duration of each stimulus, spanning 12 s. To derive the transient response in the time domain, the obtained responses underwent another round of filtering through a second-order Butterworth band-pass filter between 0.1 and 15 Hz.

## III. RESULTS

Fig. 1 displays results of CI1 obtained from two distinct sessions (Session A and Session B) conducted in separate months. The data collected in both sessions exhibit promising findings, suggesting the potential viability of employing the paradigm introduced in [1] for bilateral CI users, particularly at lower pulse rates of 40 and 80 pps. However, a comparison of these outcomes with those responses of NH listeners evoked by filtered clicks, as detailed in [1], prompts several pertinent observations and assumptions.

**Assumption 1**: In contrast to the NH results in [1], the larger offset responses shown in both session A and B might stem from the increased contamination of CI stimulation artefacts in the offset responses compared to the onset responses.

**Assumption 2**: While the eASSR at 40 pps falls within a similar range as NH results, the eASSR evoked by the 160 pps biphasic pulse train raises scepticism. Its amplitude is approximately twice that of both the 40 and 80 pps pulse trains. This is in contrast to the young NH group, where the ASSR evoked by filtered clicks adheres to the order of 40-Hz-ASSR > 160-Hz-ASSR > 80-Hz-ASSR > 320-Hz-ASSR. This suggests that components that are excessively and not in line with NH responses are likely to be contaminated by robust CI stimulation artefacts.

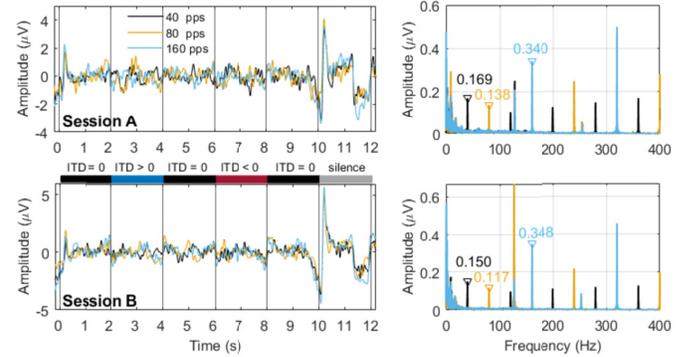

**Fig. 1** The average responses presented in both the time domain (left panels) and frequency domain (right panels). The black, yellow, and blue colours correspond to pulse rates of 40, 80, and 160 pps.

**Assumption 3**: Unexpectedly, strong responses around 127 Hz and additional harmonic frequencies were observed in all the stimuli with different pulse rates. The origins of these frequency components remain unclear. One plausible explanation could be introduced by certain hardware within the recording pipeline, such as potentially the automatically inserted power-up frame processing.

In summary, the outcomes of CI1's data present promising results regarding the feasibility of recording multiple responses from bilateral CI users using a similar paradigm as proposed in [1]. However, the aforementioned questions require careful consideration when interpreting the findings and it is necessary to explore potential optimizations of the setup.

**A. Validation of assumption 1**

To investigate potential causes for the larger offset responses compared to the onset responses, an example averaged raw EEG recording for the 80 pps stimuli is illustrated in the top panel of Fig. 2. As depicted in Fig. 2, the artefacts manifest at distinct switching time points: onset, four ITD switches, and offset. This information can be employed to check the stimuli used in the experiment and meticulously adjust the trigger time, as demonstrated in Fig. 2. Moreover, by plotting the EEG recordings contaminated with CI artefacts, users gain insight into the final stimuli presented to participants, enabling further optimization of the setup if necessary. For instance, from an inspection of Fig. 2, two potential optimizations in the current setup can be identified.

**Power Up Frames Adjustment:** Automated insertion of power up frames occurs both before switching on and after switching off stimuli for Cochlear CI device. However, the time delays differ, approximately 300 ms before switching on and around 110 ms after switching off. This discrepancy can contribute to larger artefacts in offset responses than in onset responses, arising from the overlap between offset responses, and CI power-off artefacts. A plausible solution involves manually incorporating longer power up frames (e.g., > 400 ms) before stimulus onset and after offset.

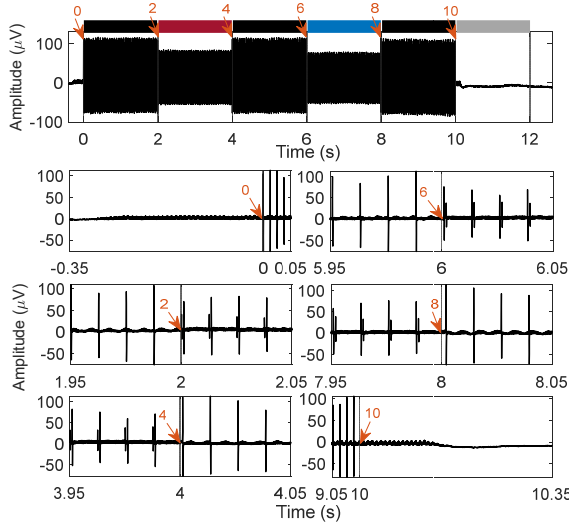

**Fig. 2** The averaged responses of the 30 segments of raw EEG data for the 80 pps stimuli. The lower panels are zoomed-in plots around 0, 2, 4, 6, 8, and 10 seconds (red arrows with numbers).

**CAEPs Latency Compensation:** Upon examination of the zoomed-in plots around 2, 4, 6, and 8 s, it becomes evident that the anticipated ITD value (e.g., 1000 μs in Fig. 2) is consistent. Nevertheless, a systematic increasing shift along the time scale is observable due to cumulative errors introduced in the system. This issue can be addressed by either optimizing the stimuli generation or compensating for the corresponding shift derived from artefact-contaminated raw data when calculating the latency of various CAEPs.

**B. Validation of assumption 2 and 3**
In our NH study [1], we demonstrated that wavelet-based time-frequency visualization can provide additional insights compared to analysing the time or frequency domain alone. Fig. 3 presents time-frequency scalograms generated using an adaptive super-resolution wavelet transform [31] for responses elicited by pulse trains of 40, 80, and 160 pps, respectively. The colour scales indicate the power distribution, expressed in arbitrary units. To emphasize the time-frequency region enclosed by the purple rectangle and to enhance the power contrast among distinct responses, the lower panels depict the corresponding plots using a more compressed scale.

Within these scalograms, there is noticeable energy concentration around 127 Hz, particularly coinciding with the start and end of the power up frames detailed in Fig. 2. This observation lends additional support to our hypothesis that the frequency component around 127 Hz might be linked to the automatic insertion of power up frames within the Cochlear CI device. To validate this proposition more thoroughly, it would be necessary to gather data from a larger cohort of Cochlear CI users (switching on/off power up frames), and users of other CI brands.

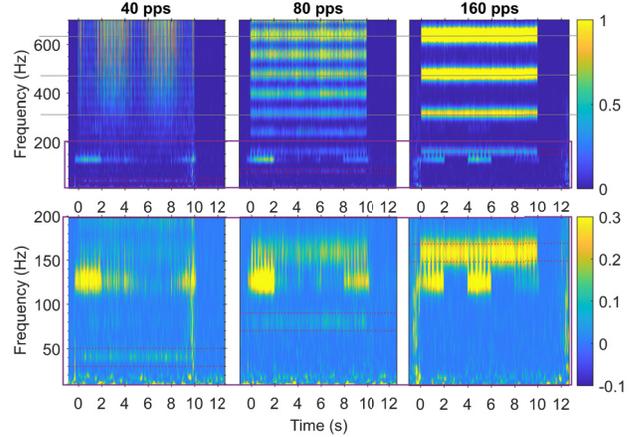

**Fig. 3** The time-frequency scalograms generated using an adaptive super-resolution wavelet transform for responses elicited by pulse trains at 40, 80, and 160 pps, respectively.

In summary, the insights from Fig. 2 and Fig. 3 highlight potential benefits of CI stimulation artefacts for refinement in the experimental setup and stimuli control.

## IV. DISCUSSION AND CONCLUSIONS

We collected EEG data from a bilateral Cochlear CI user for different pulse rates to show how CI stimulation artefacts can be used to improve experimental setups, especially for bilateral CI stimulation. We highlighted some insights and possible improvements for the current setup. For example, the CI artefacts were employed to meticulously adjust the trigger time in the exemplary data. We might be able to reduce the interference of streamer artefacts on the onset and offset responses by adjusting the power-up frames that are automatically inserted. We also recommend using shorter stimulus duration to minimize cumulative errors introduced by the streamer over time. For instance, we plan to shorten the 12-s stimulus to 6 s by testing only onset, one ITD switching ACC, and offset in the future. Moreover, for the 40 and 80 pps data, we only need to apply band-pass filters to remove CI artefacts, which is convenient for clinical applications. However, for pulse rates above 80 pps, we found a strong unknown artefact around 127 Hz in the current system that requires more advanced methods to eliminate. Both the power up streamer artefacts and the robust artefacts around 127 Hz were not detected by a standard oscilloscope with a CI detector box setup during our initial check, which further supports that exploring the CI artefacts can have some benefits for experimental design and for developing research methods. Time-frequency analysis, such as super-resolution wavelet transforms, can assist the identification of potential cochlear implant artefacts.